\documentclass[preprint]{aastex}

\shorttitle{NGC 2168 (M35) and NGC 2323 (M50)}
\shortauthors{Kalirai, J. S. et al.}

\begin{document}


\title{The CFHT Open Star Cluster Survey. IV. Two Rich, Young 
Open Star Clusters: NGC 2168 (M35) and NGC 2323 (M50)}


\author{Jasonjot Singh Kalirai\altaffilmark{1,2}}


\author{Gregory G. Fahlman\altaffilmark{3}}
\author{Harvey B. Richer\altaffilmark{1}}
\author{Paolo Ventura\altaffilmark{4}}


\altaffiltext{1}{Department of Physics \& Astronomy, 6224
Agricultural Road, University of British Columbia, Vancouver BC, Canada, 
V6T 1Z1; jkalirai@physics.ubc.ca, richer@astro.ubc.ca}
\altaffiltext{2}{Visiting Astronomer, Canada-France-Hawaii Telescope operated 
by the National Research Council of Canada, the Center National de la Recherche 
Scientifique de France and the University of Hawaii}
\altaffiltext{3}{National Research Council of Canada, Herzberg Institute of 
Astrophysics, 5071 West Saanich Road, RR5, Victoria BC, Canada, V9E 2E7; 
Greg.Fahlman@nrc.gc.ca}
\altaffiltext{4}{Osservatorio Astronomico di Roma, Sede di
Monteporzio Catone, Via Frascati 33, I-00040, Italy; 
paolo@coma.mporzio.astro.it}


\begin{abstract}
We continue our study of rich Galactic clusters by presenting 
deep CCD observations of both NGC 2168 (M35) and NGC 2323 (M50).  
Both clusters are found to be rich (NGC 2168 contains at least 
1000 stars brighter than $V$ = 22 and NGC 2323 contains $\sim$2100 stars 
brighter than our photometric limit of $V$ $\sim$ 23) and young 
(age of NGC 2168 = 180 Myrs, age of NGC 2323 = 130 Myrs).  
The color-magnitude diagrams for the clusters exhibit clear main 
sequences stretching over 14 magnitudes in the $V$, 
$B{\rm-}V$ plane.  Comparing these long main sequences with those of 
earlier clusters in the survey, as well as with the Hyades, 
has allowed for accurate distances to be established 
for each cluster ($d$ of NGC 2168 = 912 $\pm \ ^{70}_{65}$ pc, $d$ of 
NGC 2323 = 1000 $\pm \ ^{81}_{75}$ pc).  
Analysis of the luminosity and mass functions suggest 
that despite their young ages, both clusters are somewhat dynamically 
relaxed exhibiting signs of mass-segregation.  This is especially 
interesting in the case of NGC 2323, which has an age of only 1.3 times 
the dynamical relaxation time. The present photometry is also deep 
enough to detect all of the white dwarfs in both clusters.  We discuss 
some interesting candidates which may be the remnants of quite 
massive ($M$ $\geq$ 5$M_\odot$) progenitor stars.  The white dwarf cooling 
age of NGC 2168 is found to be in good agreement with the main-sequence 
turn-off age.  These objects are potentially very important for setting 
constraints on the white dwarf initial-final mass relationship and 
upper mass limit for white dwarf production.

\end{abstract}


\section{Introduction} \label{intro}

The CFHT Open Star Cluster Survey (Kalirai et al. 2001a, 
hereafter Paper I) is a large imaging program primarily designed to 
catalogue hundreds of white dwarf (WD) stars in open star clusters.  
Spectroscopic observations of these WDs with 8-m class 
telescopes (currently underway) will yield important observational 
constraints on both the WD initial-final mass relationship 
and the upper mass limit to WD production.  In addition to studying 
WDs,  the survey has motivated several other 
investigations.  For example, we are using a grid of new  
theoretical isochrones, calculated especially for this project, to 
determine both the ages of the clusters and to confront the 
theoretical predictions with observational data. With accurate photometry 
down to $V$ = 24 for the richest clusters, the quality of the data 
has made these tests much more reliable than in the past.  The long, tightly 
constrained main sequences of the clusters are also being fit to 
the Hyades and Pleiades main sequences to determine accurate distances.  
Dynamical effects are being investigated by using updated 
mass-luminosity relations to convert luminosity functions into mass 
functions.  The wide-field photometry for individual clusters has also allowed 
us to study the cluster extent, the stellar density profiles, mass-segregation 
and binary star counts.  Other studies concerned with proper motions, variable stars, 
radial velocities, brown dwarfs, synthetic CMD tests, blue stragglers, binary white 
dwarfs and Galactic disk star distributions, are currently being undertaken through 
collaborations.   Detailed results on the first cluster we examined, NGC 6819, are 
published in Kalirai et al. 2001b, hereafter Paper II.  The second cluster we looked 
at, NGC 2099, is described in Kalirai et al. 2001c, hereafter Paper III.   This work, 
Paper IV in our series, presents the study of the two young clusters NGC 2168 (M35) 
and NGC 2323 (M50).

Our deep $V$,$B{\rm-}V$ photometry covers a larger areal extent for each cluster 
than previous studies, presents the tightest main sequence to date and extends 
the photometry to bluer stars and therefore detects the faintest white dwarfs.  
Our primary goals in this work are to better constrain the ages and distances to 
each cluster, investigate the cluster luminosity and mass functions for dynamical 
effects and catalogue potential white dwarfs for future spectroscopic observations.

Perhaps the most important motivation for the study of young open star clusters is 
to set constraints on the initial mass function (IMF).  Unlike the general 
field population, clusters provide samples of stars which contain the same metallicity 
and are of the same age.  For young clusters where dynamics has not yet altered 
the stellar population, the present day observed mass function is  a close approximation 
to the IMF. NGC 2168 and NGC 2323 are good candidates for these studies.

NGC 2168 (M35) is a well-studied cluster as its rich stellar population has long 
been recognized \citet{smart,trumpler,ebbighausen,meurers}.  The most comprehensive 
of these early studies, that of \citet{ebbighausen}, detected 150 stars down to a 
photovisual magnitude of 12.0.  This study presented both the first CMD (actually 
plotting $m_{bol}$ vs log($T_{eff}$)) and luminosity function for NGC 2168.  
Thirty years later, \citet{cudworth} used photographic plates dating back to 
1916 to measure proper motions of 763 stars in the region of NGC 2168 (513 
confirmed cluster members).  He extended the cluster CMD down to $V$ $\sim$ 14.  

Preliminary dynamical studies of stars in the region of NGC 2168 were motivated by 
Galactic structure studies. The density gradients near the anti-center of the 
Galaxy were shown to be significantly steeper than other directions in the 
Solar neighborhood \citep{hersperger,topaktas}, and therefore they indicated 
anisotropies in the disk of the Milky Way. 

NGC 2168 is an anti-center cluster in the plane of the disk $(l^{II} = 
186^{\rm o}58^{\prime}, b^{II} = 2^{\rm o}18^{\prime})$ (1950), and thus was mapped 
out in these three color (RGU) studies.  However, the first significant study of 
dynamics within NGC 2168 did not occur until 1986 when \citet{mcNamara} 
investigated velocity dispersions in the cluster. They observed mass-segregation 
and proposed a model for dynamical evolution in the cluster.   In a separate 
effort, \citet{mcNamara1} also extended the CMD down to $V$ $\sim$ 15 and improved 
the photometric precision and calibration over previous studies (which were based 
on the photoelectric work of \citet{hoag}).  The dynamical mass within the central 
3.75 pc of NGC 2168 was estimated by \citet{leonard} to be in the range of 1600 
to 3200 $M_\odot$.  This measurement used two-component velocity dispersion 
profiles (from the 61 year epoch difference between photographic plates) 
to estimate the cluster potential.   

The first CCD photometry for NGC 2168 were those of \citet{sung}, 
who determined the age, reddening, distance, metallicity and minimum 
binary fraction.  Their data extended the cluster photometry to 
$V$ $\sim$ 20.  Even more recently, the WIYN Open Star Cluster 
Study has looked at both the metallicity \citep{deliyannis,deliyannis1} 
and the CMD \citep{vonHippel2,vonHippel} for NGC 2168 and drawn comparisons 
between this cluster and the Pleiades.  Finally, \citet{barrado1} 
have presented a very detailed investigation of stars from close to the turn-off 
to the hydrogen burning limit in NGC 2168.  

With so many studies of its population, the key cluster parameters of NGC 2168 
(reddening, distance and age) have been refined many times in the literature as 
observations have improved.  We summarize these in Table 1.  The best-fit 
values are those from the recent work of \citet{barrado}.  
Unlike for NGC 2168, previous work on NGC 2323 (M50) has been very limited.  
Although, several independent investigations of the cluster exist, they 
are mostly limited to observations of only a few stars.   This is surprising 
considering that the cluster is smaller in angular size, richer in stellar 
population and younger in age than NGC 2168.  Early photographic and photoelectric 
studies of NGC 2323 found distances which varied by over a factor of two: 
\citet{trumpler} - 780$\rm -$860 pc, \citet{shapley} - 500$\rm -$800 pc, 
\citet{collinder} - 675 pc, \citet{rieke} - 520 pc, \citet{cuffey} - 1210 pc, 
\citet{hoag} - 1170 pc, \citet{mostafa} - 995 pc.  The first age determinations 
for the cluster varied by similar factors: \citet{barbaro} - 60 Myrs, 
\citet{mostafa} - 140 Myrs.  More recently, \citet{claria} presented $UBV$ 
photoelectric measurements of 109 probable members of the cluster.  This study, 
which has a limiting magnitude of $V$ $\sim$ 14, established $E(B{\rm-}V)$ = 0.25, 
$(m{\rm-}M)_{V}$ = 10.62 ($d$ = 940 pc), and age = 100 $\pm$ 20 Myrs 
(from fitting \citet{bertelli} isochrones with mass loss and moderate 
overshooting).  No previous CCD studies of this cluster were found 
in the literature.  The present photometry detects $\sim$10$\times$ as 
many cluster members as all previous studies combined.

We will first present the data and briefly discuss the photometry 
and calibration in \S 2.  In \S\S 3 and 4 we present the CMDs for each 
cluster and fit isochrones to the data.  This allows us to determine 
the age of each cluster and also investigate the quality of the fits.  
In \S 5 we discuss dynamical effects in both clusters and compute the 
luminosity and mass functions.  Finally, we present analysis of white 
dwarfs in NGC 2168 and NGC 2323 in \S 6, and conclude this study in \S 7.

\section{Observations and Data Reduction}\label{observations}

As with other clusters in the CFHT Open Star Cluster Survey, the data were obtained 
during a three night observing run in 1999 October. All observations were obtained
at the Canada-France-Hawaii Telescope through standard $V$ and $B$ filters with the 
CFH12K mosaic camera, which has a 42$'$ $\times$ 28$'$ field-of-view.  NGC 2168 was 
imaged during night 3 of the run, and NGC 2323 during night 1 (short exposures to 
get the photometry of 
giants were acquired at a later date).  For NGC 2168, two 5-minute exposures were 
obtained in each color, however, only one was usable.  This was combined with individual 
50, 10 and 1 second exposures to complete the photometry.  For NGC 2323, the photometry 
from single exposures of 300, 50, 10 and 1 seconds in each color was combined to 
produce a catalogue of magnitudes and colors.  The seeing conditions during the 
observations of NGC 2168 were 1.15$^{\prime\prime}$ and the airmass was 1.60.  
For NGC 2323, the seeing was 0.85$^{\prime\prime}$ and the airmass was 1.18.  

The data were processed (flat-field, bias and dark corrected) and combined using 
the FITS Large Images Processing Software (FLIPS) (J.-C. Cuillandre 2001, private 
communication) as described in \S 3 of Paper I.  Over 53 flats and 13 bias frames 
were observed during the run.  

We reduced the data using a preliminary version of the TERAPIX photometry routine 
PSFex (Point Spread Function Extractor) (E. Bertin  2000, private communication).  
We used a separate, variable PSF for each CCD in the mosaic.  Some 
features of PSFex are described in \S 4 of Paper I.  

Calibration was addressed by separately observing Landolt standard 
fields SA-92 and SA-95 \citep{landolt} during the observing run.  A total of 92 
calibration stars (different configurations of exposure times, filters and 
airmasses) were used to calibrate the NGC 2168 data.  For NGC 2323, 130 calibration 
stars were used (see Tables 2 and 3 in Paper I).  Comparison of the aperture 
photometry of these stars with the PSFex photometry yielded the transformation 
equations required to calibrate the data (see \S\S 5.1 and 5.2 of Paper I).  The  
uncertainty in the zero points  used for the calibration is estimated to be $\sim$0.018 
in $V$ and $\sim$0.017 in $B$. The air-mass coefficients are 0.087 $\pm$ 0.012 in $V$ and 
0.160 $\pm$ 0.006 in $B$, in good agreement with CFHT estimations of 0.10 and 0.17 
respectively. The color terms were averaged over the three night observing run and 
are in agreement with the nominal CFHT estimates in the $V$ filter and are slightly 
lower for the $B$ filter.

The final errors in the photometry are found to be less than 0.01 magnitudes for 
stars with V $<$ 19, and up to 0.2 magnitudes for the faintest (V $\sim$ 23) 
stars.

\section{The NGC 2168 and NGC 2323 Color-Magnitude Diagrams}\label{cmd}

Figures \ref{N2168_cmds} and \ref{N2323_cmds} present photometric 
color-magnitude diagrams (CMDs) for both NGC 2168 and NGC 2323.  The 
panels on the left show all objects measured with PSFex, whereas the panels on the 
right depict only those objects with a stellarity index $\geq$ 0.50.  This 
SExtractor image morphology parameter \citep{bertin} uses a robust procedure 
to estimate source resolution and we have shown in the past (Papers II and III) 
that this cut is effective in removing faint-blue galaxies and other image defects.  
Clearly, the fainter parts of the CMDs in the right panels are much cleaner than 
the left panels.  Some potential white dwarf candidates are also brought out in 
the faint blue end of the CMDs.

The CMD for NGC 2168 (Figure \ref{N2168_cmds}) shows a rich main sequence extending 
from $V$ $\sim$ 8.5 to $V$ $\sim$ 22.5.  One potential cluster red giant is also 
seen at $V$ = 8.48, $B{\rm-}V$ = 1.32 and can be used to constrain the age of the 
cluster through fitting of the blue loop.  Although it is centrally located in 
the cluster, this object is only given a 56\% probability as a cluster member in 
the proper-motion analysis by \citet{mcNamara}.  The main sequence presented here is 
tighter than any of the previous studies on this cluster.  This photometry is 
deep enough to detect the faintest white dwarfs in the cluster.  These are believed 
to have originated from quite massive progenitors.  The `clump' of objects near 
$V$ $\sim$ 21.5, $B{\rm-}V$ $\sim$ 0 most likely represents the end of the white 
dwarf cooling sequence in this young cluster.  This will be further 
investigated in \S \ref{wds}.  The wall of stars bluer (or fainter) than the cluster 
main sequence is the background disk star population.  With a Galactic latitude of 
only $\sim$2 degrees, NGC 2168 resides very close to the plane of the Milky Way.  
Within our 40$^{\prime}$ diameter for the cluster field (one direction is cutoff at 
28$^{\prime}$ by the CFH12K mosaic) we have detected many field stars which overlap 
with cluster members on the faint main sequence.  The vertical feature in the CMD to 
the red of the cluster main sequence (at 14 $\leq$ $V$ $\leq$ 17.5, $B{\rm-}V$ $\sim$ 
1.3) may represent distant giant stars in the disk.  This is smeared out in $V$ due to a 
distance spread along the line of sight.  The  defined field does not encompass 
the entire radial extent of NGC 2168.

For NGC 2323, the CMD (Figure \ref{N2323_cmds}) again shows a rich, long main 
sequence extending from the tip of the turnoff at $V$ $\sim$ 8 down to $V$ $\sim$ 
23 at the limit of our data.  As mentioned in \S \ref{intro}, the previous best 
study of this cluster, the photoelectric effort of \citet{claria}, only detected 
109 cluster members brighter than $V$ $\sim$ 14.  The SExtractor cleaned CMD 
(right panel) again shows several potential white dwarfs scattered in 
the faint, blue end of the diagram.

\subsection{Cluster Reddening, Distance and Metallicity}\label{parameters}

As summarised in Table 1, several previous measurements of both 
the reddening and distance of each of NGC 2168 and NGC 2323 are available 
in the literature.  We adopt reddening values from these studies, and 
re-compute the distances by comparing the cluster main sequences to the 
Hyades fiducial \citep{deBruijne}.  Since there is minimal scatter in 
the main sequences for these clusters, there is almost no ambiguity in 
the distance determination.  For NGC 2168, we 
adopt $E(B{\rm-}V)$ = 0.20 from the recent 5-color ($UBVRI$) photometry of 
\citet{sarrazine}.  The metallicity of NGC 2168 is known due to the high 
resolution spectroscopic effort of \cite{barrado} who find $[Fe/H]$ = 
${\rm-}$0.21 $\pm$ 0.10 ($Z$ $\sim$ 0.01).  Other previous CCD studies, (e.g., 
Sung \& Bessell 1999) have also found sub-Solar metallicities for NGC 
2168 using less robust techniques.  To compare the Hyades sequence to 
NGC 2168, we first shift our CMD by +0.09 magnitudes in $B{\rm-}V$ to account 
for the $\Delta$$Z$ $\sim$ 0.014 metallicity difference (Hyades is given as 
$Z$ = 0.024).  The corresponding best fit distance modulus is found to be 
$(m{\rm-}M)_{\rm 0}$ = 9.80 $\pm$ 0.16, which gives d = 912 $\pm \ ^{70}_{65}$ 
pc.  The extinction has been corrected by using $A_{V}$ = 3.1$E(B{\rm-}V)$.  
This distance estimate is slightly greater than previous values for NGC 2168 
(see Table 1), however, it is more precise 
because of the quality of the CMD.

For NGC 2323, we adopt a reddening value of $E(B{\rm-}V)$ = 0.22, which is an 
average of the recent \citet{claria} study and the robust effort of 
\cite{hoag}.  Although no detailed spectroscopic abundance analysis is 
published for NGC 2323, photometric investigations (e.g., Claria, Piatti 
\& Lapasset 1998) have found that Solar metallicity models best reproduce 
the observed cluster main sequence.  We shift the CMD by +0.03 magnitudes 
in $B{\rm-}V$ to account for the $\Delta$Z $\sim$ 0.004 metallicity 
difference and find that the best fit distance modulus for NGC 2323 is 
$(m{\rm-}M)_{\rm 0}$ = 10.00 $\pm$ 0.17, which gives d = 1000 $\pm \ ^{81}_{75}$ 
pc.  As Table 1 shows, this distance is in good agreement with the most recent 
studies of NGC 2323

The uncertainties in the distances for these clusters are computed using 
an identical method to that used for NGC 2099 in Paper III.  Briefly, the 
final uncertainty in the true distance modulus combines four individual 
uncertainties, and accounts for correlations between them.  These are (1) 
a scale factor translation from the $B{\rm-}V$ axis to the $V$ axis to account for 
the reddening uncertainty, (2) an estimated main-sequence fitting uncertainty at
a fixed reddening value, (3) the error in the extinction, $\Delta_{A_{V}}$ = 
3.1$\Delta_{E(B{\rm-}V)}$ and (4) the color uncertainty due to an estimated metallicity 
uncertainty, $\Delta_{Z}$.  Paper III describes the details of how these terms 
were combined.

\section{Theoretical Comparisons}\label{theory}

As with earlier clusters in the survey, we are using a new set of 
theoretical isochrones to compare the observations to theory.  
This set of models was computed especially for the CFHT Open 
Star Cluster Survey by the group at the Rome Observatory (P. Ventura 
\& F. D'Antona). \S 5.2 of Paper II and \S 3.4 of Paper III provide 
a description of the models.  Here we present only a small 
summary and refer the readers to the earlier papers for more insight.  
A description of the stellar evolution code 
that was used to build the tracks is given in \cite{ventura}.  Convective 
core-overshooting is treated by means of an exponential decay of turbulent 
velocity out of the formal convective borders, and the convective flux has 
been evaluated according to the Full Spectrum of Turbulence (FST) theory 
prescriptions \citep{canuto}.  The \cite{bessell} conversions are used to 
convert luminosities to magnitudes, and temperatures to colors.  The lower main 
sequence ($M$ $\lesssim$ 0.7 $M_\odot$) has been calculated by adopting NextGen 
atmosphere models \citep{hauschildt}.

Figure \ref{N2168_2isos} shows two panels with the best fit isochrones for NGC 
2168.  For the core-overshooting model shown in the left panel, we find an age 
of 180 Myrs.  This is essentially identical to the 175 Myrs value found by 
\cite{barrado} and very close to the 150 Myrs value found by \citep{vonHippel2}. 
There is a potential red giant star at $M_{V}$ = -1.94, $(B{\rm-}V)_{\rm 0}$ 
= 1.21 which helps constrain the location of the blue loop.  This star (as well 
as a turn-off star located at $M_{V}$ = -1.76, $(B{\rm-}V)_{\rm 0}$ = -0.09) 
fall directly under the line which is used to draw the isochrone on the figure 
and therefore is difficult to see.  We have increased the point size of all 
stars with $M_{V} \leq$ 0 to help visualize the data.  A non-overshooting model 
is also shown on the right and this gives an age of 130 Myrs.  The $\zeta$ value in 
the figures represents the free parameter that specifies the e-folding distance of 
the exponential decay of turbulent velocities \citep{ventura}.

The isochrone reproduces the general shape of the main sequence very 
nicely.  Near the turn-off, the model correctly falls on the blue-edge 
of the observed sequence which is free of binary-contamination.  Models of 
ages 160-200 Myrs in the overshooting case also provide acceptable fits 
to the observed cluster sequence.   For the lower main sequence ($M_{V} 
\gtrsim$ 8), we were not able to compute non-grey atmospheres for this 
metallicity.  We therefore overplot the lower main sequence from a Solar 
metallicity ($Z$ = 0.02) model, and therefore a discontinuity is seen in 
the CMD.  A small color shift (of approximately $\Delta$($B{\rm-}V$) = 0.06) is 
needed to correct for the metallicity of the cluster.  Such a shift would move 
the lower main sequence blueward, but not by an amount which would produce a 
poor fit.  

Figure \ref{N2323_2isos} shows isochrone comparisons for NGC 2323.  The age, 
using a core-overshooting model is found to be 120-140 Myrs (left panel).  The 
Solar metallicity isochrone fits the observed main sequence and turn-off well.  As 
in the case of the similar metallicity clusters NGC 6819 and NGC 2099, the lower 
main sequence in this model again falls slightly bluer than the data.  This result 
is, however, a great improvement over previous grey atmosphere isochrones.

\section{Selection of Cluster Members}\label{selection}

\subsection{Previous Estimations and Control Fields}\label{control}

The most comprehensive study of NGC 2168's population \citep{barrado1} 
used main sequence counting techniques to discover $\sim$1700 cluster members 
in their data.  When combined with data from other studies (brighter stars), 
the total cluster mass, based on an average mass of 0.6 $M_\odot$, is estimated 
to be $\sim$1600 $M_\odot$. \citet{sung}, used mass function integrations to also 
estimate a total mass of 1660 $M_\odot$ for the cluster, although only 640 $M_\odot$ 
of stars were counted in the study.  An earlier dynamical study by \citet{leonard} 
placed the total cluster mass anywhere between 1600-3200 $M_\odot$.  

All of these photometric studies of NGC 2168 have suffered in establishing the true 
cluster population given observational limits due to its large size.  The spatial 
extent of the cluster easily fills the areal coverage in most detectors and hence 
there are no documented faint studies of a blank field offset by at least 1 degree 
from the core.  Although there is a small effect in these studies which will 
underestimate the total cluster population because of missed stars near the edge of 
the cluster, the true population may actually be overestimated due to improper 
blank field subtraction.  To try and get around this, \citet{barrado1} used a wide 
strip centered on the main-sequence locus and counted all 
stars within this envelope.  Field star subtraction was then estimated 
by using counts of stars surrounding the main sequence.  The filter choices 
chosen by \citet{barrado1} separates the disk stars from the cluster stars better 
than a $V$, $B{\rm-}V$ CMD, however, their cluster luminosity function may 
still include some faint K and M-type disk dwarfs.

In the study of \citet{sung}, the observed cluster luminosity function was 
extrapolated beyond $V$ $\sim$ 17.5 (mass function beyond $\sim$0.7 $M_\odot$) based 
on counts in the Pleiades cluster.  However, as mentioned in their paper, 
this overestimates the contribution of low-mass stars as the Pleiades sample 
includes the low-mass dominated outer regions of the cluster, which are not 
detected by \citet{sung}.  For the brighter stars, they used proper motion 
selection (a better technique of getting cluster membership), however for the 
fainter stars they adjusted ZAMS lines in different filters for reddening 
excesses and computed distances for individual objects.  Objects with inconsistent 
distances are assumed to be disk stars and thrown out.  

In the work of \citet{leonard}, an age of 30 Myrs was used to represent NGC 
2168 (from the work of \citet{vidal}), much lower than the value found in recent 
studies.  Additionally, most of the mass in the \citet{leonard} study resided in 
faint main-sequence stars which were beyond the proper motion detections.  Therefore, 
an extrapolation was used to integrate the MF using a power-law slope beyond 
$\sim$1 $M_\odot$.  Binary effects and mass-segregation were ignored in 
the study

Our study also suffers from a small enough field of view that a 1:1 cluster field 
equivalent blank field (in terms of areal coverage) cannot be established.  However, 
rather than count stars over our entire mosaic we have chosen to build a blank field 
from the extreme outer regions of our field (an area of 122 ${\sq^{\prime}}$, 
outside of R = 20$^\prime$ from the center of the cluster).  This 
field is almost a factor of 8 smaller in areal coverage than our cluster field.  
At this radius, the CMD shows no evidence of a main sequence however some cluster 
stars are most likely still present as previous proper motion studies have shown 
(see \S \ref{intro}).  Current generation instruments, such as MEGACAM on CFHT, 
will easily allow a deep blank field to be obtained around this cluster.  

For NGC 2323, the only study investigating the luminosity function 
of the cluster is the $V$ $\leq$ 14.5 effort of \citet{claria}.  They 
concluded with a lower limit on the cluster population of 109 members 
(285 $M_\odot$).  In this case, our areal coverage is not only large enough 
to encompass almost the entire cluster ($R \sim$ 15$^{\prime}$), but also 
large enough so that an almost areal equivalent blank field can be build up 
from the outer CCDs on the mosaic.

\subsection{Luminosity and Mass Functions}\label{lumfunc}

We define the cluster stars by first creating main-sequence fiducials 
(defined by clipping objects with $\Delta(B{\rm-}V)$ $\geq$ 3.5$\sigma$ from 
the mean) after roughly isolating the cluster main sequence from the background 
distribution.  This  creates an envelope around the fiducial based on the errors 
in the photometry (the envelope broadens out towards faint magnitudes) and 
stars are counted within this envelope, for both the cluster CMD and the 
background CMD.  The observed cluster luminosity function is the difference 
between the counts in the two fields after accounting for the difference in area  
between the cluster and background fields.

Incompleteness corrections are handled by repeatedly introducing a number 
of artificial stars into the data images.  The method used for this has been 
described in both Papers II and III, so we only present the results here.  Table 2 
shows the statistics for these tests in both the cluster and blank fields 
(ratio of number added to number recovered) in column 4.  The incompleteness 
tests were carried out for the NGC 2168 data set and the recovered results are 
used for both clusters.  This method is believed to be reliable since both of 
these clusters were imaged for identical exposure times and exhibit similar crowding.  
The corrections found here are very similar to those determined for NGC 
2099 in Paper III.  To summarize: the completeness in the cluster field for main-sequence 
stars is found to be 73.5\% at $V$ = 22.5.

The final corrected star counts are found by multiplying the cluster and blank field 
counts by the incompleteness corrections from Table 2.  The final luminosity functions 
for each cluster are tabulated in columns 2 and 3 of Table 2 (corrected for 
incompleteness).  Figures \ref{N2168_lum} and \ref{N2323_lum} show the corresponding 
luminosity functions for NGC 2168 (solid line) and NGC 2323.  For NGC 2168, the 
luminosity function rises until $M_{V}$ = 5, and then dips down and rises 
again.  Although this dip ($M_{V} \sim$ 7) is most likely the result of poor 
statistics or cluster stars being subtracted off in the blank field and not 
physical, we do not rule out the latter.  Many open cluster CMDs, such as all 
four that we have looked at in detail in our survey, appear to show fewer stars 
in a thin band of the lower main sequence near this magnitude.  We do note that 
the luminosity function given in \citet{barrado1} does not show a dip at this 
magnitude (dashed line).  This was constructed by using the $I$ band luminosity 
function in Table 3 of 
\cite{barrado1} and correcting to $V$ by using the empirical fiducial in Table 1 of 
that paper.   For NGC 2323, the luminosity function shows a steady climb to the 
limit of our data where there is a small dip.  Again, the last data 
point is likely affected by poor statistics.  Both luminosity functions show a slow 
rise beyond $M_{V} \sim$ 8.  This rise is due to the change of slope in the 
mass-luminosity relation \citep[see e.g.,][]{dantona}, and may be seen as a change in 
slope of the cluster main sequence.

By summing the corrected luminosity function, we find a cluster population 
of just over 1000 stars down to $V$ $\sim$ 22 in the central 20$^\prime$ of 
NGC 2168.  The number of stars here has not been corrected for the chopped off 
area on the N and S edges of the mosaic.  Due to this, the effective area of this 
field is actually equivalent to a spherical 18$^\prime$ field.  Since our background 
field may contain some cluster stars, the above estimate is likely to be a 
lower limit.  This number increases significantly when the cluster luminosity function 
is extended to the hydrogen burning limit.  For this, we normalize our luminosity 
function with respect to the \cite{barrado1} function, and predict that there are 
$\sim$500 additional stars between $V$ = 22-25.

For NGC 2323, we detect $\sim$2050 stars in the central 15$^\prime$ and down 
to $V$ $\sim$ 22.  Again, since our data does not extend to the lowest mass stars, 
we can use the well known Pleiades luminosity function \citep{lee} to estimate 
the total population down to $M_{V}$ = 15.  For this, we find that an 
additional 1150 faint stars would exist in the cluster, raising the total 
population to 3200.

Figure \ref{LumFuncComp} compares the luminosity functions for both 
NGC 2168 and NGC 2323 to that for the Solar neighbourhood \citep{binney2}, 
the Pleiades \citep{lee}, and the 520 Myr old rich cluster NGC 2099 
\citep{kalirai1}.  Each of the functions have been normalized by the 
total number of stars and then shifted arbitrarily in the vertical 
direction.  The clusters are presented in order of increasing age, 
with the Pleiades (the youngest) at the top of the diagram, directly 
beneath the Solar neighborhood luminosity function.  It is 
clear that this data does not reach deep enough to see the turn-over in 
the luminosity functions of NGC 2168 or NGC 2323 (the last data point in 
each function has been eliminated given the large error bar). 

To compute the mass functions of these clusters, we multiply the observed 
luminosity functions by the slope of the mass-luminosity relationship (from the 
models).  The resulting mass function is usually expressed as a power law, 

\begin{equation}
\Psi(m) \propto m^{-(1+x)}, \label{salpeter}
\end{equation}

\noindent where $x$ = 1.35 represents the Salpeter value.  For NGC 6819, 
we found a very flat mass function ($x$ = $\rm-$0.15).  This was expected as 
the cluster is very old (10$\times$ the dynamical relaxation time).  For NGC 
2099, the best-fit slope was found to be $x$ = 0.60.  Other studies, specifically 
\citet{francic} have shown that older clusters systematically show flatter mass 
functions. This is believed to be due to the preferential loss of low mass stars, 
a consequence of dynamical evolution toward energy equiparition.  

The mass functions of NGC 2168 and NGC 2323 are presented in Figures \ref{N2168_massfunc} 
and \ref{N2323_massfunc}.  Fitting a slope to the global NGC 2168 mass function gives a 
value of $x$ = 1.29 $\pm$ 0.27, almost identical to the Salpeter slope.  Our result 
splits the values found in previous studies;  \cite{leonard} found $x$ = 1.68 for the 
extreme high mass end of the cluster mass function,  \cite{sung} found a flatter value 
of $x$ = 1.1, and \cite{barrado1} found $x$ = 1.59 for 0.8-6 $M_{\odot}$.  For 
NGC 2323, we find $x$ = 1.94 $\pm$ 0.15 over the mass range (0.40-3.90 $M_\odot$).  
No previous estimations exist in the literature.  Therefore, for the four 
clusters that we have yet looked at in the CFHT Open Star Cluster Survey, 
the mass function is found to be systematically steeper as the cluster age 
decreases.

\subsubsection{Dynamical State and Mass-segregation}\label{dynamics}

	As mentioned earlier, young open star clusters are excellent 
test cases to determine the IMF as they represent a group of stars 
with a similar age and composition.  However, this IMF can only be 
determined if the cluster has retained the stellar population that it 
was born with; that population has not suffered from preferential 
mass loss.  This mass loss occurs as the cluster relaxes toward a 
state of energy equipartition due to distant two-body interactions 
between all the stars within the cluster.  In an equilibrium state, the lower 
mass stars will be moving with a higher velocity and therefore will be more 
likely found in the outskirts of the cluster. There, they are preferentially 
lost from the cluster tidal boundary. Thus dynamical relaxation leads 
to mass-segregation, manifested by a systematic change in the slope of the local 
mass function with radius, and to an evolution of the global cluster IMF, 
generally from a relatively steep slope (high value of $x$) to a smaller slope, 
with time.  If we can show that there is no evidence of this dynamically 
induced mass-segregation, then there is hope in setting constraints on 
the IMF.  If rather, we find that dynamical effects have shifted the stellar 
distribution, then these dynamical effects can still lead to an important 
understanding of the timescales and levels of mass-segregation in clusters.  
The caveat is that we must disentangle dynamical evolution from 
primordial mass-segregation, a process that may be inherent to cloud collapse 
to begin with.

\citet{binney} characterize the dynamical relaxation time in terms of 
the number of crossings of a star that are required for its velocity 
to change by order of itself, 

\begin{equation}
t_{relax} \sim t_{cross}\frac{N}{8lnN}, \label{eqnrelaxationtime}
\end{equation}
\medskip

\noindent where $N$ is the total number of stars.  Using our distance and 
apparent size of NGC 2168, we estimate that the dynamical relaxation age is 
111 Myrs, which implies that the cluster age is 1.6$\times$ its relaxation age.  
For NGC 2323, the dynamical age is 102 Myrs, which gives a cluster age of less 
than 1.3$\times$ the relaxation age. 

To investigate signs of mass-segregation, we split the clusters into four 
annuli (with geometry given in Table 3), and present mass functions in each 
annulus in Figures \ref{N2168_massseg} and \ref{N2323_massseg}.  Note that 
each annulus with a radius greater than $\sim$14$^\prime$ is cut off horizontally 
at the top and bottom due to the edges of CFH12K (which are 14.06$^{\prime}$ from 
the center of the mosaic).  The results show that mass-segregation is most likely 
present in both clusters.  The outer annuli have steeper mass functions, 
indicating more lower mass stars relative to higher mass ones.  Surprisingly, 
the effect is much more prominent in the younger NGC 2323, for which 
the mass function slope goes from $x$ = 2.74 $\pm$ 0.40 in the outer annuli 
(11$^{\prime} \leq$ $R$ $<$ 15$^{\prime}$) to $x$ = 0.82 $\pm$ 0.27 in 
the innermost annuli (0$^{\prime} \leq$ $R$ $<$ 3$^{\prime}$).  For NGC 
2168, the mass function slope changes from $x$ = 1.48 $\pm$ 0.96 in the 
outer annuli (15$^{\prime} \leq$ $R$ $<$ 20$^{\prime}$) to $x$ = 0.66 
$\pm$ 0.32 in the innermost annuli (0$^{\prime} \leq$ $R$ $<$ 
5$^{\prime}$). 

As mentioned earlier, the mild mass-segregation that we see in these 
young clusters could be primordial.  Different clusters themselves may form 
with different concentrations of high mass stars in the core, and low mass 
stars in the halo.  This type of initial mass-segregation has been observed in 
extremely young open clusters \citep{zinnecker} where the clusters are younger 
than the estimated relaxation timescales.  Formation scenarios for these young 
clusters stress that the rapidity and efficiency of star formation is crucial in 
determining the final distribution of the stars \citep{battinelli}.  Supernovae 
and stellar winds from the first generation of stars can effectively blow 
away remaining gas in the system to make the cluster unstable.  The 
subsequent dynamical evolution of the cluster will be affected by the 
timescales and stages of this gas removal \citep{stecklum}.  As for the IMF, 
it too depends on the local conditions of cloud collapse.  For 
example, the final slope of the IMF will depend on how many cloudlets 
formed protostellar cores through collisions before the gravitational 
influence of the collapse ceased to be important \citep{murray}.  Since 
the collisional frequency of the cloudlets will be greatest in 
higher density regions, such as the centers of clusters, more massive 
stars will also preferentially form in the center.  This leads to a 
natural mass-segregation within the cluster.

We can resolve these two cases of mass-segregation induced either from dynamical 
evolution or primordial mass-segregation by looking for signs of energy 
equipartition.  Generally, dynamical mass-segregation can be demonstrated 
if stars of different masses obey the radial distributions predicted by 
multi-mass King models \citep{gunn}.  This will be investigated for a series 
of clusters in a forth-coming paper.

\section{White Dwarfs}\label{wds}

The global goal of our survey is to first identify white dwarf 
candidates in open star clusters and then obtain spectra to determine their 
masses.  This will help us understand both the upper mass limit to white 
dwarf production and the initial-final mass relationship which is a critical 
input for calculating ages from white dwarf cooling.  The upper mass limit is 
also important in chemical evolution models of galaxies as it sets the lower 
mass limit for type II supernova.  White dwarfs in NGC 2168 
and NGC 2323 represent the most important sample in this study as they 
constrain the high mass end of the relationship.  With core-overshooting 
main-sequence turn-off masses of $\sim$5 $M_\odot$, some white dwarfs in 
both of these clusters are expected to have massive progenitors.

Our photometry for these clusters is deep enough to detect the faintest 
white dwarfs which could have cooled during the lifetime of the clusters.  
Based on the cluster ages, we expect the faintest white dwarfs (assuming 
1.0 $M_\odot$ objects) in both clusters to be located at $V \sim$ 22.  The 
younger age of NGC 2323 pushes this limit to a brighter magnitude, 
however the larger distance modulus has an opposing effect.  The 
combination of small number statistics in the faint-blue end of the 
CMDs and much smaller blank fields do not allow us to build 
white dwarf luminosity functions.  We will however attempt an estimation 
of the field white dwarf numbers by scaling up the blank field numbers 
to the area of the cluster field, and also by using the Galactic disk 
white dwarf luminosity functions \citep{leggett}.  
Previously, the only white dwarfs spectroscopically confirmed in either of 
these clusters are the two objects measured by \citet{reimers} in NGC 2168.  
We believe further spectroscopy of new candidates in the present photometry 
of this cluster will yield more objects.  The cleaned CMDs (removal of 
galaxies and image defects, see \S \ref{cmd}) do show some 
concentrations of objects.  For example, a clump of objects can 
be seen in NGC 2168 between 21 $\leq$ $V$ $\leq$ 22.5, $B{\rm-}V$ = 0.  
For NGC 2323 more of a scattered distribution of objects is seen in 
the white dwarf region.

We can predict the number of expected white dwarfs in each of these 
clusters by integrating the main-sequence mass functions from the 
present day turn-off mass to the upper limit for white dwarf 
production.    For the latter, our models show $M_{\rm up}$ 
$\sim$7 $M_\odot$, however the value could be as low as 5.5 $M_\odot$ 
\citep{jeffries}.  The calculation of the number of expected 
WDs assumes that the observed mass function shape is similar to that 
for the higher masses which have evolved off the main sequence.  Given 
both the range of masses represented by the present day mass functions 
(NGC 2323 -- 0.4 $M_\odot$ to 3.9 $M_\odot$), and the smoothness of the 
data, only a mild extrapolation to higher masses is needed for these 
young clusters.  For NGC 2168, we use the mass function slope from 0.83 
$M_\odot$ to 3.56 $M_\odot$ ($x$ = 1.75) to estimate 18 expected white 
dwarfs in our cluster field.  For NGC 2323, the mass function slope from 0.79 
$M_\odot$ to 3.90 $M_\odot$ ($x$ = 2.42) predicts 6 white dwarfs in the 
cluster.  

Before we can compare the number of white dwarfs seen to the number 
expected, we must first remove the number of field white dwarfs 
(assuming galaxies have already been removed through the stellarity 
cut).  For this, we use the disk white dwarf luminosity function 
\citep{leggett} and scale the number to our field of view.  Since 
the Galactic disk is much older than these two clusters, we must 
truncate the counts in the field at the bright magnitudes which are 
represented by the white dwarf cooling limits of the clusters.  For 
the NGC 2168 field, the disk white dwarf luminosity function predicts 
4 white dwarfs down to $V$ = 22.  For NGC 2323, 3 field white dwarfs 
are predicted in the cluster field of view down to $V$ = 22 (12 are 
predicted down to the limiting magnitude).  The number of objects seen 
in the blank field after scaling up the counts given the smaller field 
of views, are greater than these numbers.  Counting objects on the CMD, 
and eliminating these blank field numbers gives 14.5 WDs in NGC 2168 
after accounting for incompleteness errors.  This number is in excellent 
agreement with the 18 estimated from the mass function extrapolation.  
For NGC 2323, 5 white dwarfs are found after the statistical 
subtraction and 6 were predicted.  Although we do not know which 
of the objects on our CMD are cluster white dwarfs, we now have 
candidates for multi-object spectroscopic observations in these two 
clusters.  

In Figure \ref{N2168_wd}, we show a 1 $M_\odot$ cooling model 
\citep{wood} with respect to the clump of potential white dwarfs 
in NGC 2168.  The best white dwarf candidates are the 8 objects 
clustered together in the range 21 $\leq$ $V$ $\leq$ 22, -0.15 
$\leq$ $B{\rm-}V$ $\leq$ 0.1.  These are all found to be slightly 
fainter than this cooling sequence, suggesting perhaps an even 
higher mass and therefore a potentially very important constraint on 
the white dwarf initial-final mass relationship and upper mass limit 
to production.  For the 1 $M_\odot$ model, the corresponding white 
dwarf cooling ages for these objects range 
from 30 to 190 Myrs.   These values are in excellent agreement 
with the expected white dwarf cooling ages from stars which 
have evolved off the main sequence in this young cluster.  Therefore, 
if the clump of objects in NGC 2168 are bona fide cluster white dwarfs, 
then we can conclude that the white dwarf cooling age of the cluster 
(190 Myrs) is in good agreement with the main-sequence turn-off 
age (180 Myrs).  For the former age, we would also need to add on 
$\sim$50 Myrs for the main-sequence lifetime of a progenitor 
7 $M_\odot$ star before making any comparisons.


\section{Conclusion} \label{conclusion}

Deep photometry of NGC 2168 and NGC 2323 in the $V$ and $B$ filters has 
allowed us to produce the tightest main sequences to date for these 
clusters.  NGC 2168 is found to contain $\sim$1000 stars  
brighter than $V$ = 22, beyond which we can not reliably trust the luminosity 
function.  In NGC 2323 we find $\sim$2050 stars above the photometric limit 
of $V$ $\sim$ 23.  After accounting for uncertainties in 
the reddening, main-sequence fitting, extinction and metallicity, the 
distance to NGC 2168 is determined to be 912 $\pm \ ^{70}_{65}$ pc.  For 
NGC 2323, we find $d$ = 1000 $\pm \ ^{81}_{75}$ pc.  By comparing a new 
set of theoretical isochrones to the observational CMDs, we have determined both 
the ages of the clusters (180 Myrs for NGC 2168, 130 Myrs for NGC 2323) 
and also find a nice agreement between the theoretical isochrone and 
the observational main sequence.  Both clusters are 
found to exhibit mild signs of mass-segregation.  The global mass 
function of NGC 2168 is found to have a slope of $x$ = 1.29, almost 
identical to a Salpeter IMF ($x$ = 1.35).  For NGC 2323, we find $x$ = 1.94, 
steeper than the Salpeter value. Spectroscopic confirmation of potential 
white dwarfs in both clusters are important for setting high mass 
constraints on the initial-final mass relationship.  A summary of overall results 
for both NGC 2168 and NGC 2323 are given in Tables 4 and 5, respectively.


\acknowledgments


JSK received financial support during this work through NSERC 
PGS-A and PGS-B research grants.  HBR and GGF are supported in part by 
the Natural Sciences and Engineering Research Council of Canada. HBR 
extends his appreciation to the Killam Foundation and the Canada Council 
for the award of a Canada Council Killam Fellowship.


\clearpage



\clearpage

\figcaption[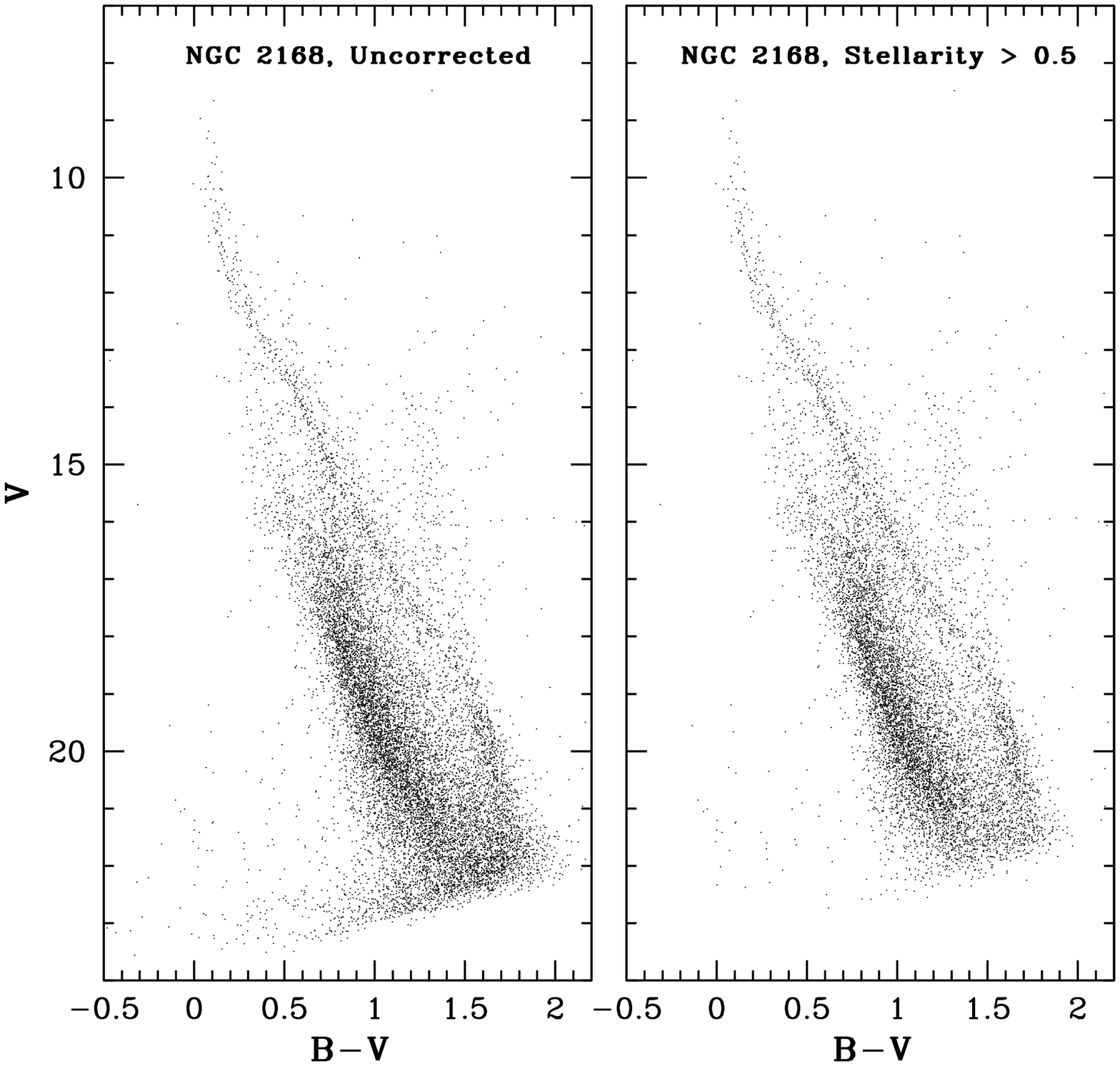]{NGC 2168 CMD clearly shows a well populated 
main sequence stretching from $V$ = 9 to $V$ $\sim$ 22.5.  After eliminating 
faint, blue galaxies and other image defects with a 0.50 stellarity cut 
(right panel), some potential white dwarfs are seen at $V$ = 21.5, $B{\rm-}V$ 
= 0. \label{N2168_cmds}}

\plotone{Kalirai.fig1.eps}

\clearpage

\figcaption[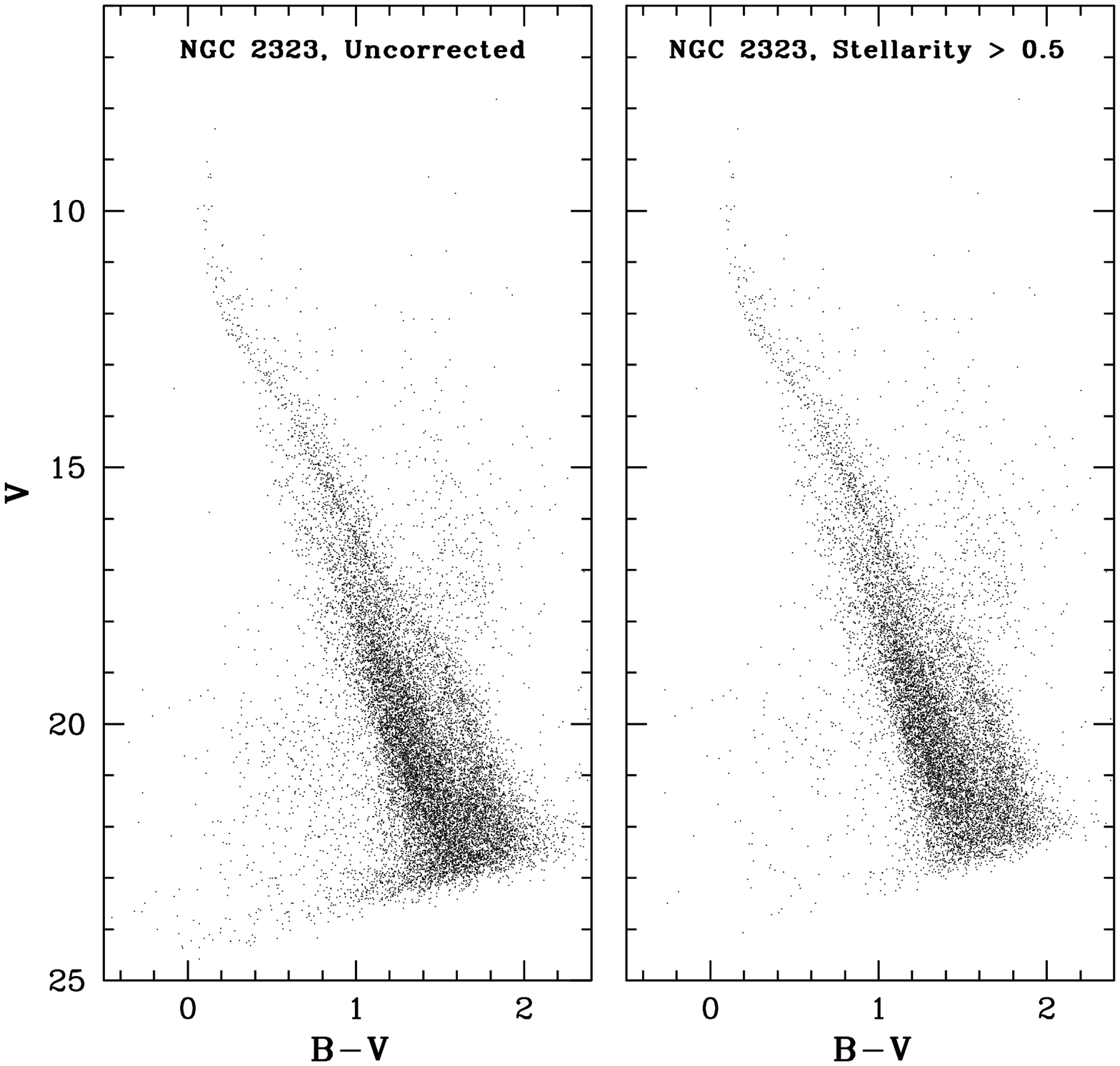]{NGC 2323 CMD clearly shows a well populated 
main sequence stretching from $V$ = 8 to $V$ $\sim$ 23.  After eliminating 
faint, blue galaxies and other image defects with a 0.50 stellarity cut 
(right panel), some potential white dwarfs are seen scattered in the 
faint-blue end of the CMD. \label{N2323_cmds}}

\plotone{Kalirai.fig2.eps}

\clearpage

\figcaption[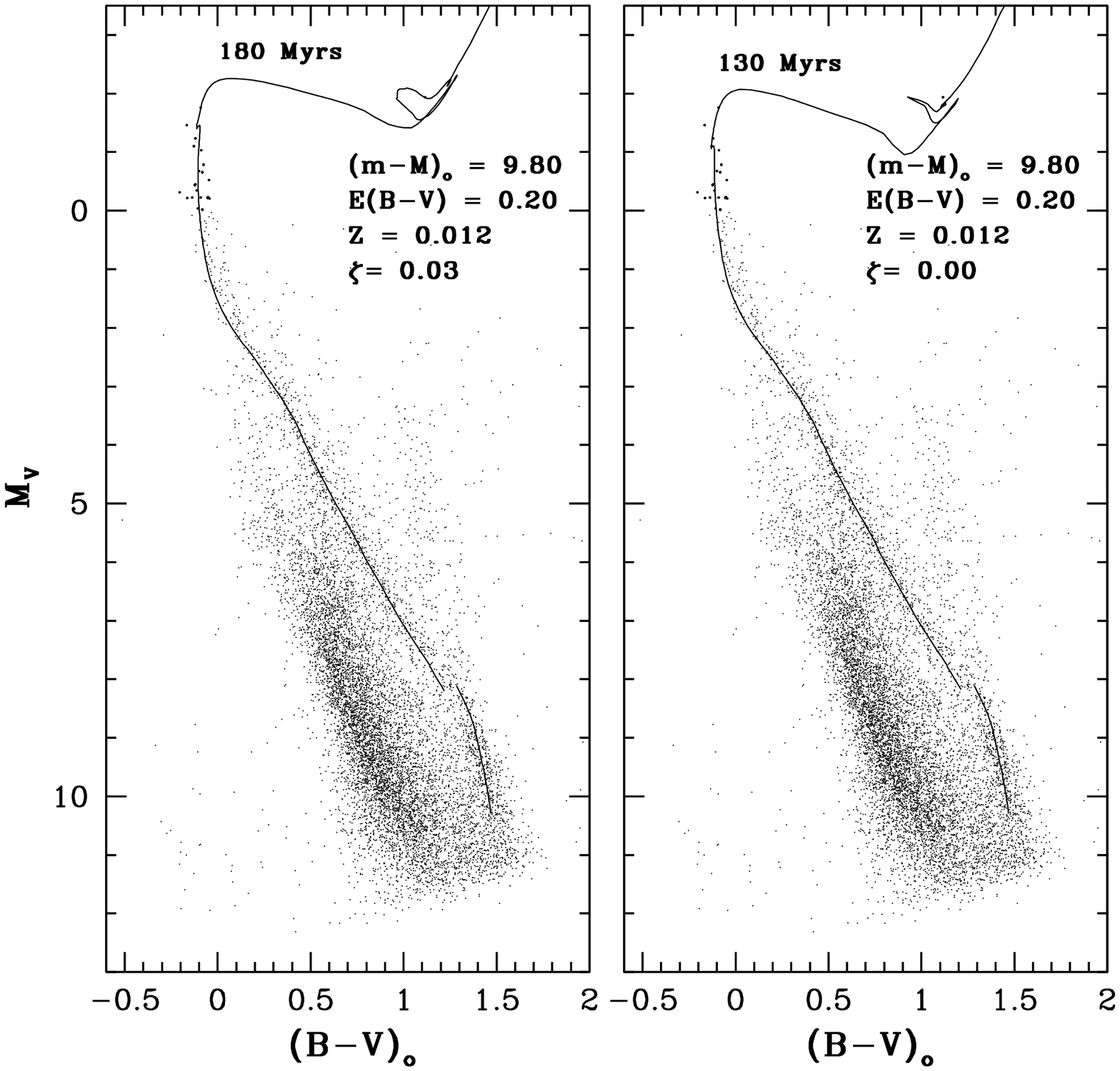]{The left panel shows the best fit
isochrones for NGC 2168 (age = 180 Myrs), based on a core-overshooting 
model.  An equivalent fit is obtained by using a non-core-overshooting 
model of 130 Myrs (right panel).  The discontinuity at $M_{V}$ = 
8.2 represents the limit to which we presently have non-grey atmosphere 
models for this metallicity ($Z$ = 0.012).  Fainter than this magnitude, 
we plot our Solar metallicity isochrones.  The metallicity difference causes 
a small shift between the colors of the two models, however, this does 
not take away from the excellent agreement on the lower main sequence 
between this model and the data.  See \S \ref{theory} for a discussion 
of these results.  Note - Dots representing stars with $M_{V}$ $\leq$ 0 have 
been made slightly larger and bolder for comparison purposes. \label{N2168_2isos}}

\epsscale{0.95}

\plotone{Kalirai.fig3.eps}

\clearpage

\figcaption[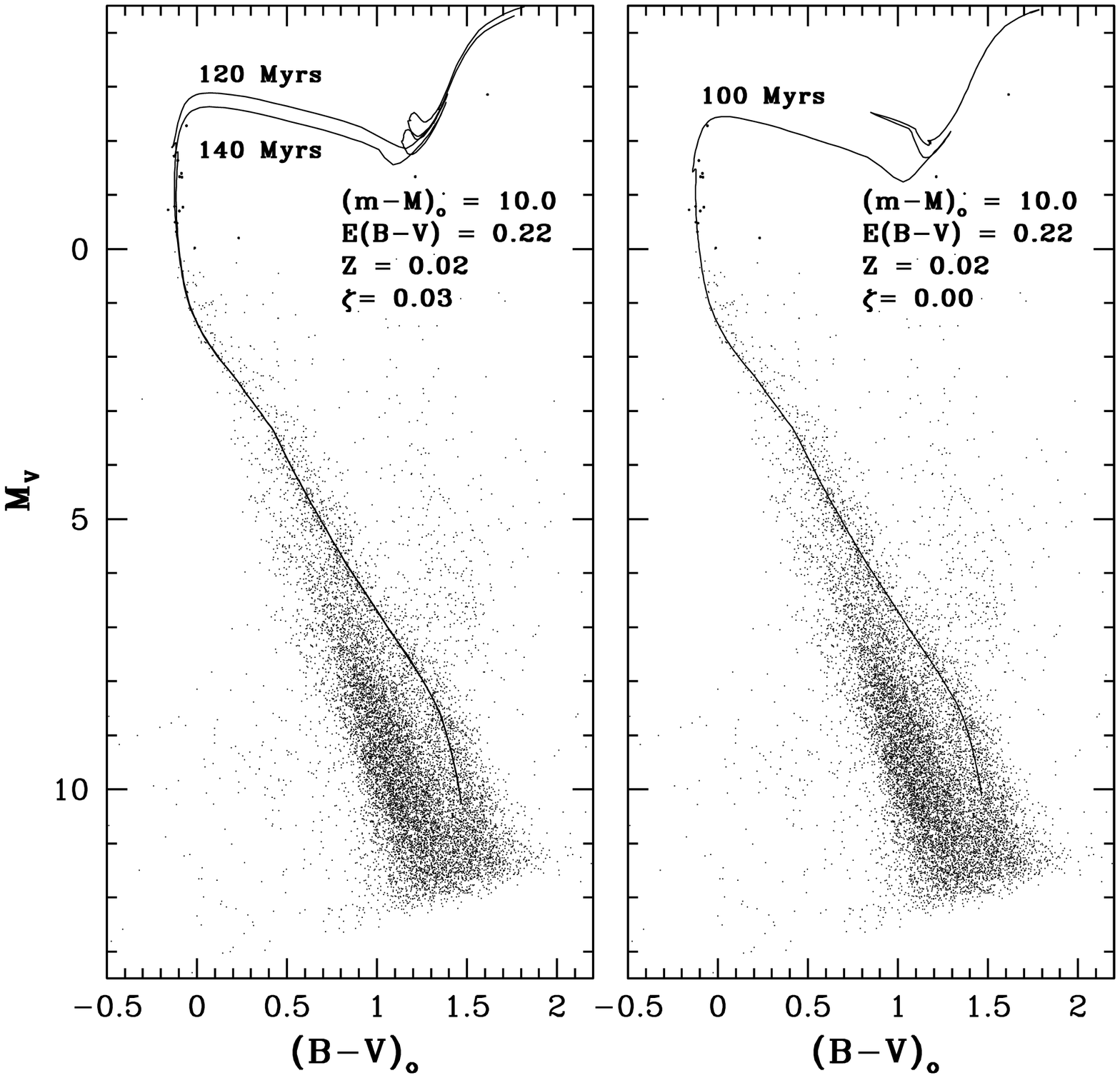]{The left panel shows the best fit
isochrones for NGC 2323 (age $\sim$ 130 Myrs), based on a core-overshooting 
model.  An equivalent fit is obtained from a non-core-overshooting 
model of 100 Myrs (right panel).  The general shape of the 
main sequence is matched well to the Solar metallicity isochrone.  
See \S \ref{theory} for a discussion of these results. \label{N2323_2isos}}

\plotone{Kalirai.fig4.eps}

\clearpage

\figcaption[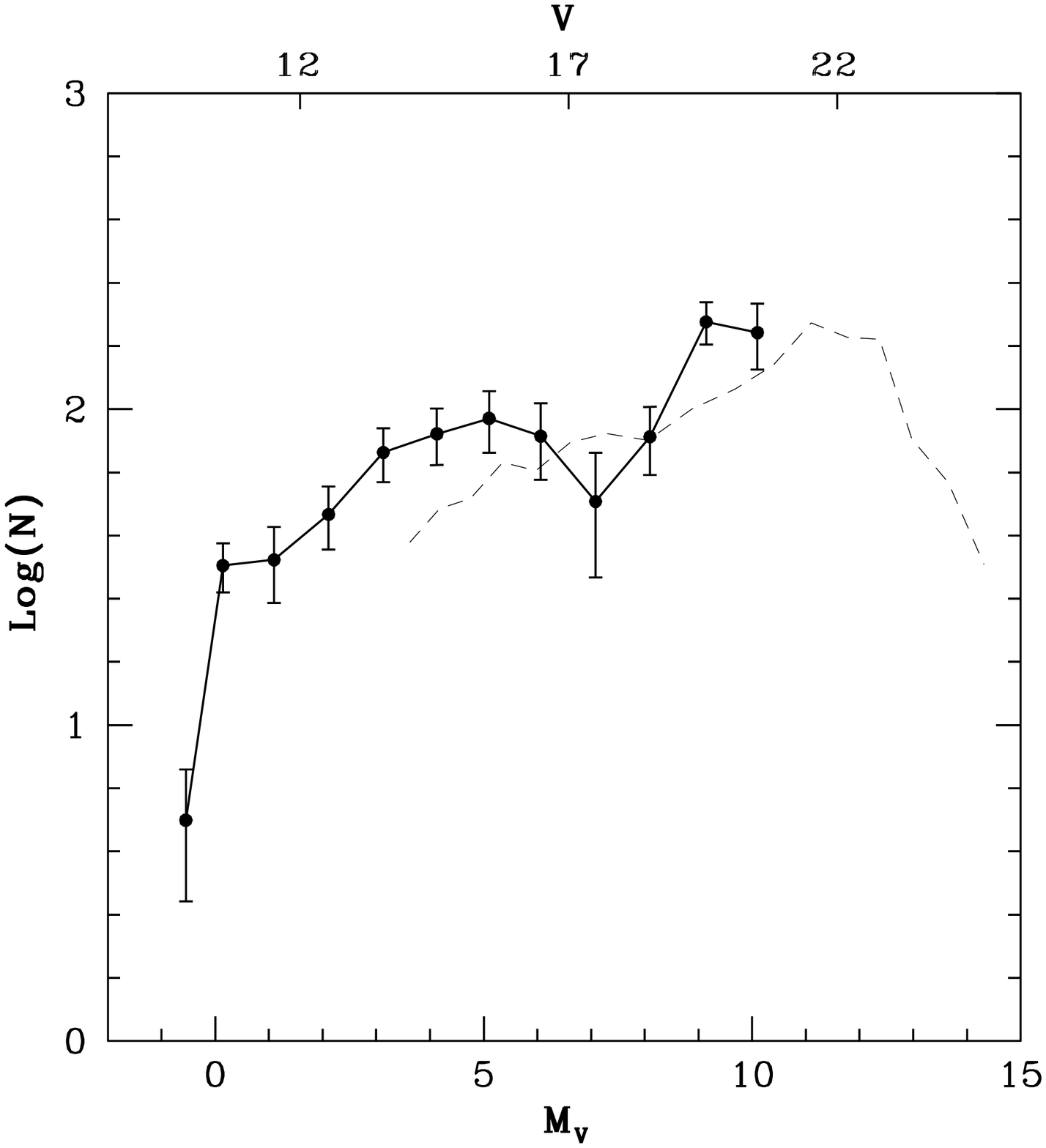]{NGC 2168 luminosity function is shown 
before after correcting for data incompleteness. The 
error bars reflect a combination of Poisson errors and incompleteness 
errors.  The NGC 2168 luminosity function from \cite{barrado1} is also shown 
for comparison.
 \label{N2168_lum}}

\plotone{Kalirai.fig5.eps}

\clearpage

\figcaption[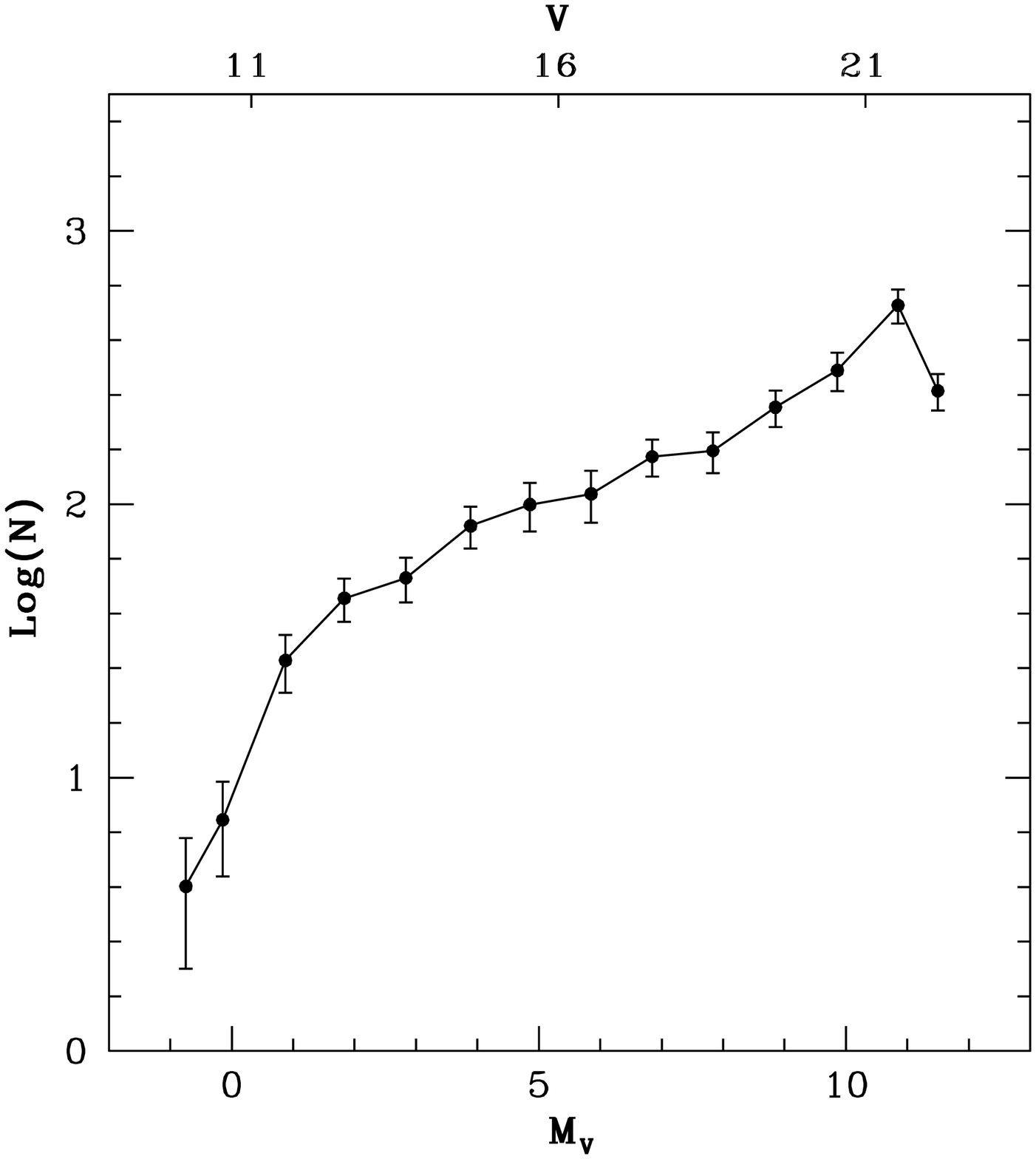]{NGC 2323 luminosity function is shown 
after correcting for incompleteness.  The 
luminosity function slowly rises up to $V \sim$ 21.5 
($M_{V}$ = 10.82).  The error bars reflect a combination of 
Poisson errors and incompleteness errors. \label{N2323_lum}}

\plotone{Kalirai.fig6.eps}

\clearpage

\figcaption[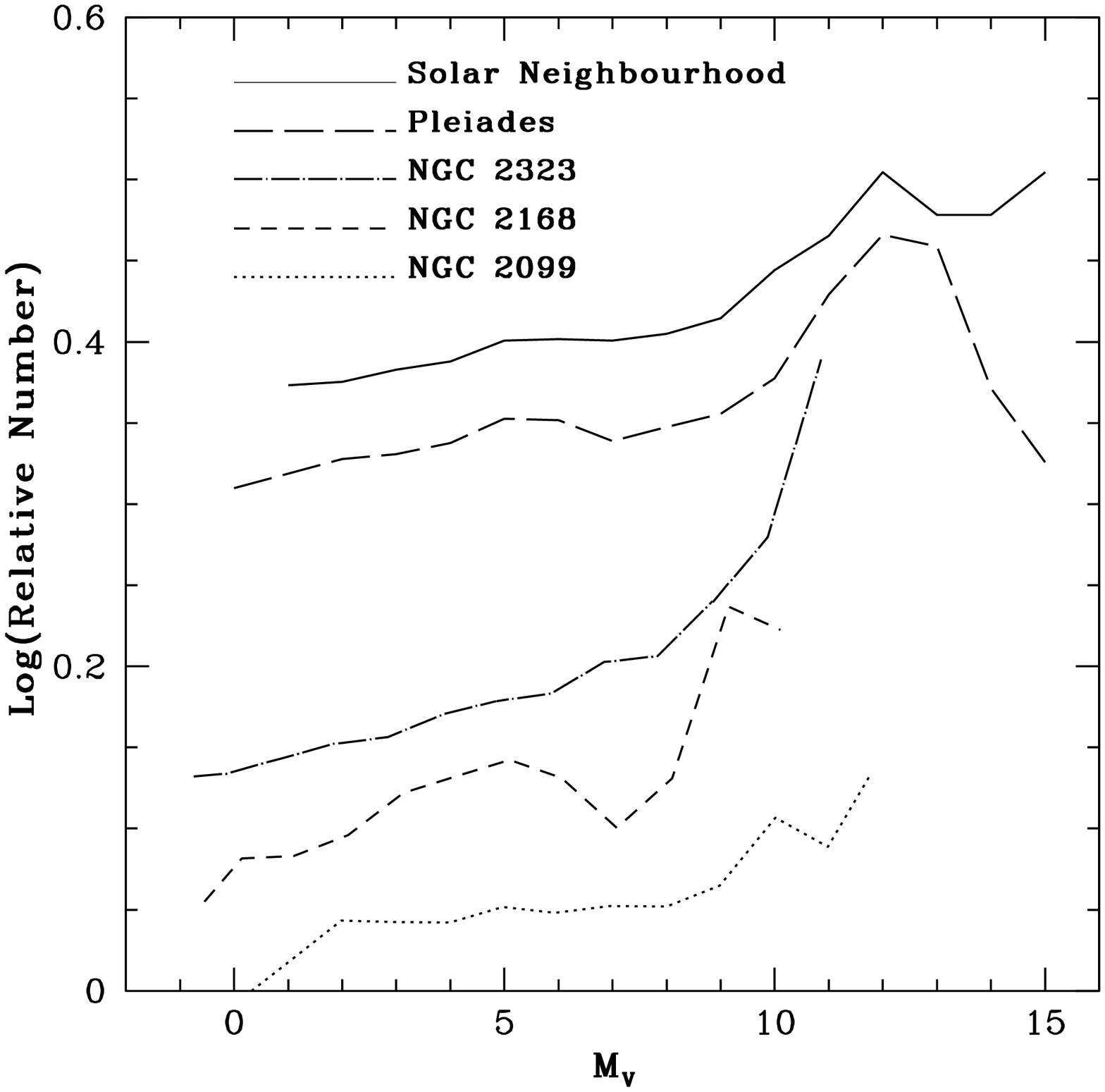]{Luminosity functions of NGC 2168 and 
NGC 2323 are compared with those of the Solar neighborhood, the 
Pleiades, and NGC 2099.  Each luminosity function has been normalized 
by the total number of stars, and then scaled up arbitrarily.  The 
clusters are shown in terms of increasing age, with the Pleiades 
being the youngest at the top under the Solar neighborhood.\label{LumFuncComp}}

\plotone{Kalirai.fig7.eps}

\clearpage

\figcaption[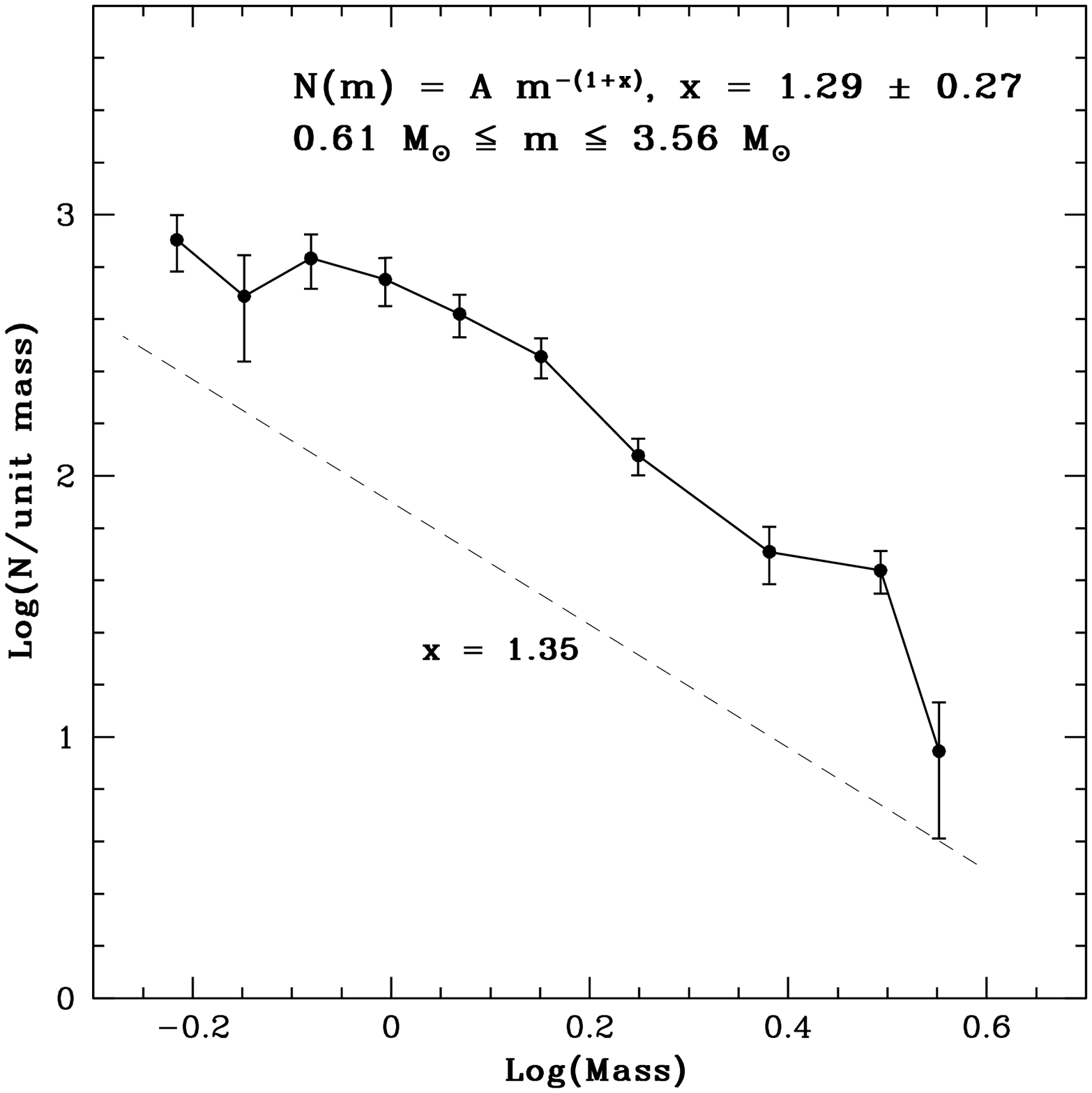]{NGC 2168 global ($R$ $<$ 20$^{\prime}$)
mass function (solid, $x$ = 1.29 $\pm$ 0.27) is found to be almost identical to 
a Salpeter IMF (dashed, $x$ = 1.35). \label{N2168_massfunc}}

\plotone{Kalirai.fig8.eps}

\clearpage

\figcaption[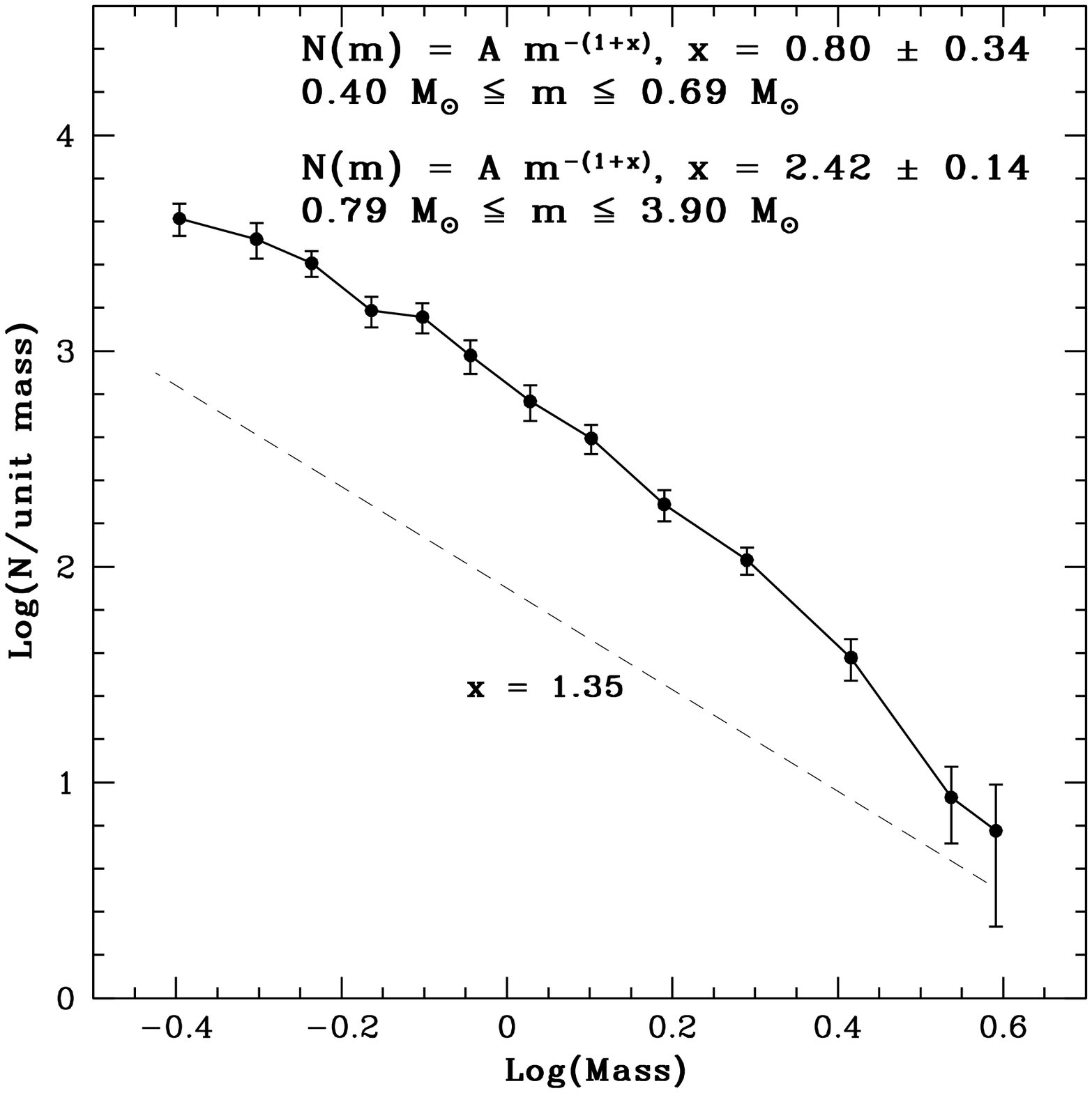]{NGC 2323 global ($R$ $<$ 15$^{\prime}$)
mass function (solid, $x$ = 1.94 $\pm$ 0.15) is found to be steeper than a 
Salpeter IMF (dashed, $x$ = 1.35). 
\label{N2323_massfunc}}

\plotone{Kalirai.fig9.eps}

\clearpage

\figcaption[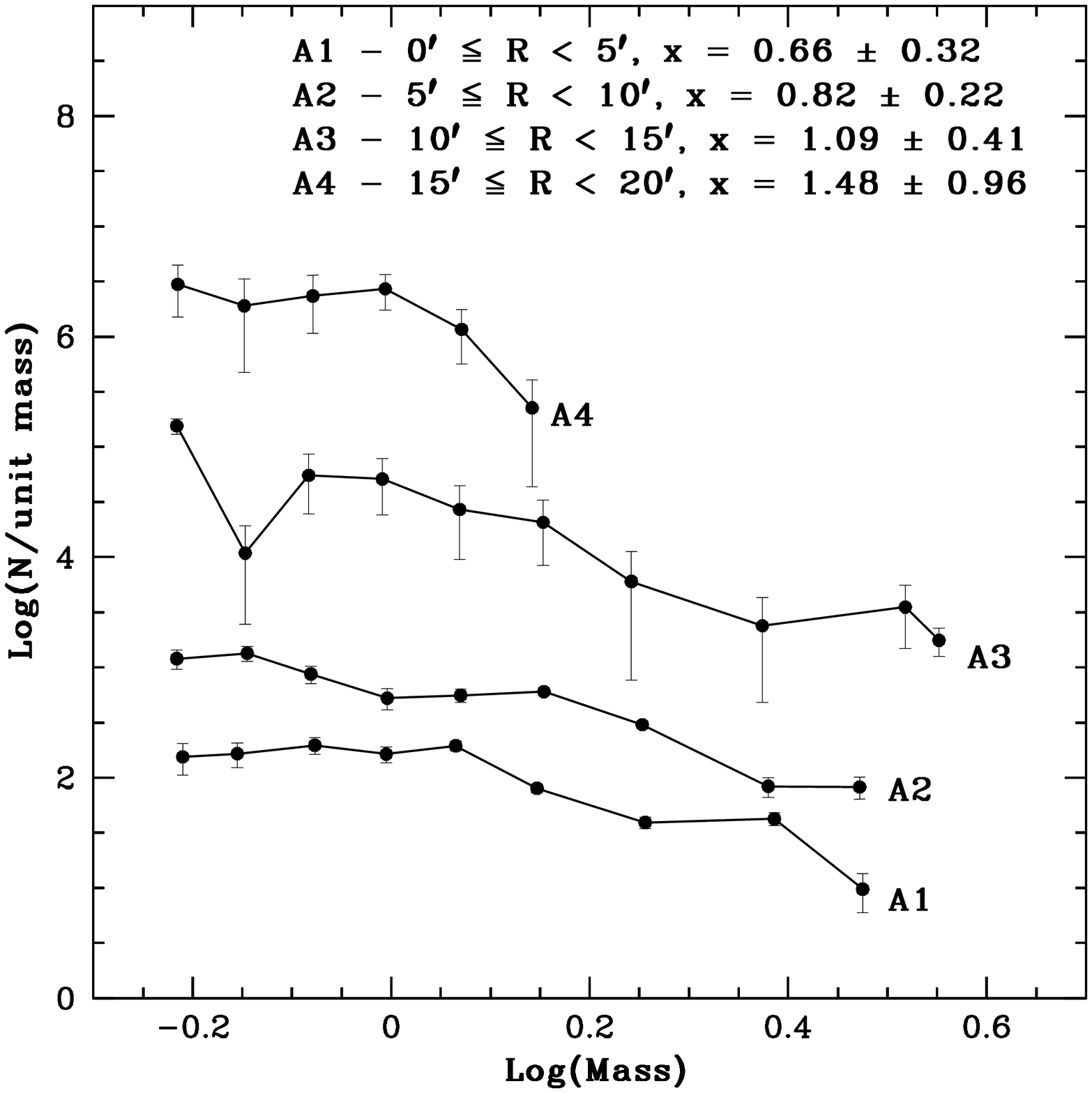]{Mass functions in different annuli 
for NGC 2168 show mild evidence for mass-segregation (steeper mass function 
in the outer rings) in the cluster. \label{N2168_massseg}}

\plotone{Kalirai.fig10.eps}

\clearpage

\figcaption[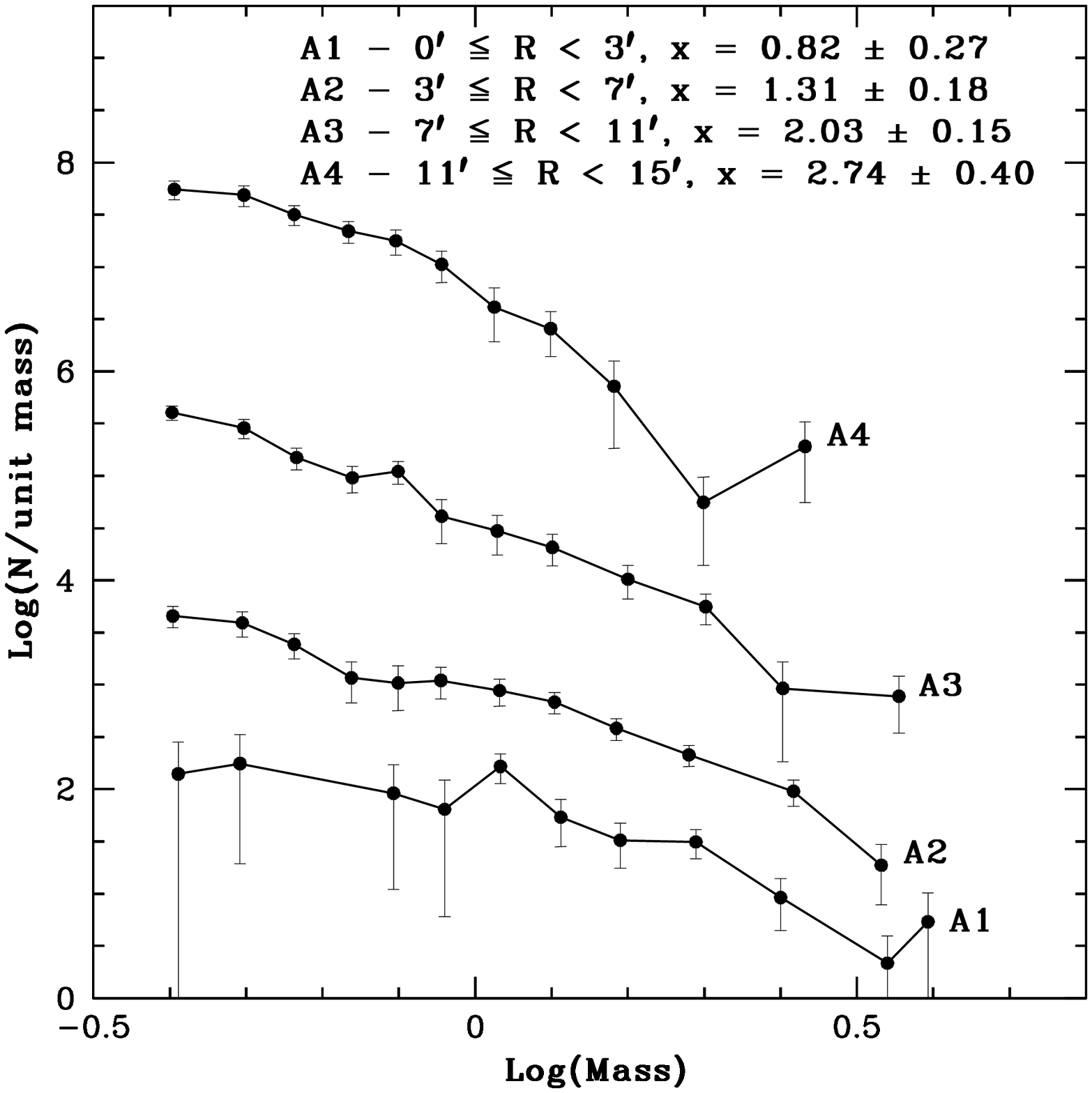]{Mass functions in different annuli 
for NGC 2323 show clear evidence for mass-segregation (steeper mass 
function in the outer rings) in the cluster despite its age being only 
1.3$\times$ the dynamical relaxation age. \label{N2323_massseg}}

\plotone{Kalirai.fig11.eps}

\clearpage

\figcaption[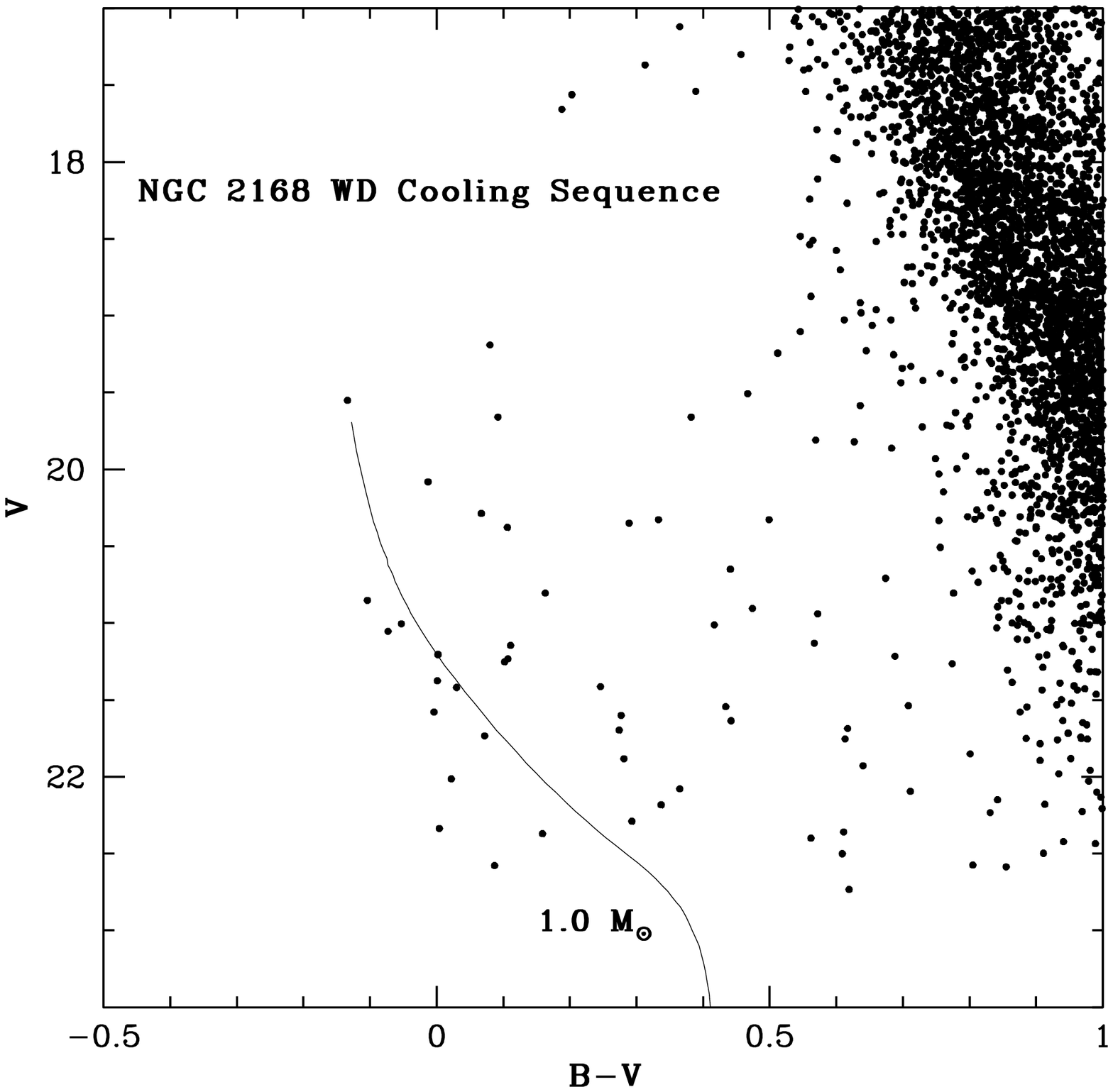]{A 1.0 $M_\odot$ white dwarf cooling 
sequence \citep{wood} is shown with respect to the potential 
white dwarfs in NGC 2168.  The clump of objects located at  
21 $\leq$ $V$ $\leq$ 22, -0.15 $\leq$ $B{\rm-}V$ $\leq$ 0.1 falls 
slightly fainter than the cooling model, indicating perhaps an 
even higher mass for these objects.  The white dwarf cooling age 
from the faintest of these stars is found to be 190 Myrs, and 
therefore in excellent agreement with the main-sequence turn-off 
age (180 Myrs). \label{N2168_wd}}

\plotone{Kalirai.fig12.eps}

\clearpage





\clearpage

\begin{deluxetable}{cccccc}
\tabletypesize{\scriptsize} \tablecaption{Summary of Published 
Cluster Parameters for NGC 2168 and NGC 2323 \label{table1}} \tablewidth{0pt}
\tablehead{\colhead{Year} & \colhead{Reddening $E(B{\rm-}V)$} &
\colhead{Distance (pc)} & \colhead{Age (Myrs)} & \colhead{Type} &
\colhead{Reference}} \startdata

\bf NGC 2168 \rm & \\

1930 & \nodata    & 840 & \nodata & Photographic     & Trumpler \\

1938 & \nodata    & 740 & \nodata & Photographic     & Cuffey \\

1961 & 0.23       & 870 & \nodata & \nodata          & Johnson et al. \\

1961 & 0.23       & 871 & \nodata & Photoelectric/Photographic & Hoag et al. \\

1971 & 0.23       & 870 & 70-100  & Photographic  & Becker \& Fenkart \\

1973 & 0.17       & 871 & 20-40   & Photographic  & Vidal \\

1992 & 0.26-0.44  & 725 & 85      & Photoelectric & Sung \& Lee \\

1999 & 0.255      & 832 & 200     & CCD      & Sung \& Bessell \\

2000 & 0.3        & 731 & 100     & CCD      & von Hippel et al. \\

2000 & 0.198      & 809 & 160     & CCD      & Sarrazine et al. \\

2001 & \nodata    & 832 & 175     & CCD      & Barrado y Navascu\'es et al. \\

2002 & 0.20       & 805 & 150     & CCD      & Deliyannix et al. \\

2003 & \nodata    & 912 & 180     & CCD      & This work \\

\bf NGC 2323 \rm & \\

1930 & \nodata   & 780-860 & \nodata  & Photographic & Trumpler \\
	      
1930 & \nodata   & 500-800 & \nodata  & Photographic & Shapley \\
	      
1931 & \nodata   & 675     & \nodata  & Photographic & Collinder \\
	      
1935 & \nodata   & 520     & \nodata  & Photographic & Rieke \\

1941 & 0.30      & 1210    & \nodata  & Photographic & Cuffey \\

1961 & 0.20-0.26 & 1170    & \nodata  & Photoelectric & Hoag et al. \\

1969 &  \nodata  & \nodata & 60       & Photographic & Barbaro, Dallaporta 
\& Fabris \\

1983 & 0.33      & 995     & 140 & Photographic & Mostafa \\

1998 & 0.25      & 931     & 100 & Photoelectric & Claria, Piatti \& Lapasset \\

2003 & \nodata   & 1000    & 130 & CCD           & This work \\

\enddata

\end{deluxetable}

\clearpage

\begin{deluxetable}{ccccccc}
\tabletypesize{\scriptsize} \tablecaption{Corrected Cluster Star Counts and 
Completeness \label{table2}} \tablewidth{0pt} \tablehead{\colhead{$V$ mag} &
\colhead{N (NGC 2168)} & \colhead{N (NGC 2323)} & \colhead{Completeness 
(Cluster/Background)}} \startdata

9.0-10.0  & 5.0 $\pm$ 2.2 & 4.0 $\pm$ 2.0  &1/1 \\

10.0-11.0 & 32.0 $\pm$ 5.7 & 7.0 $\pm$ 2.6  &1/1 \\

11.0-12.0 & 33.4 $\pm$ 9.0 & 26.8 $\pm$ 6.4 &1/1  \\

12.0-13.0 & 46.5 $\pm$ 10.5 & 45.3 $\pm$ 8.2 &1/1  \\

13.0-14.0 & 73.0 $\pm$ 14.1 & 53.7 $\pm$ 10.0 &1/1  \\

14.0-15.0 & 83.5 $\pm$ 17.0 & 83.4 $\pm$ 14.4 &1/1  \\

15.0-16.0 & 93.5 $\pm$ 20.6 & 99.6 $\pm$ 20.0 &1/1  \\

16.0-17.0 & 76.8 $\pm$ 22.3 & 109.1 $\pm$ 23.5 &1.02/1  \\

17.0-18.0 & 55.0 $\pm$ 21.7 & 149.3 $\pm$ 23.0 &1/1  \\

18.0-19.0 & 81.9 $\pm$ 19.9 & 156.8 $\pm$ 26.7 &1.03/1.02  \\

19.0-20.0 & 176.6 $\pm$ 29.0 & 226.4 $\pm$ 34.5 &1.08/1.05  \\

20.0-21.0 & 162.4 $\pm$ 41.3 & 308.7 $\pm$ 49.3 &1.13/1.09  \\

21.0-22.0 & 52.1 $\pm$ 25.7 & 534.5 $\pm$ 75.6 &1.21/1.16  \\

22.0-23.0 & \nodata & 259.7 $\pm$ 39.7 &1.36/1.18  \\

\enddata

\end{deluxetable}

\clearpage

\begin{deluxetable}{ccc}
\tabletypesize{\scriptsize} \tablecaption{Geometry of Annuli
\label{table3}} \tablewidth{0pt} \tablehead{\colhead{Annulus} &
\colhead{Radius ($'$)} & \colhead{Radius (pixels)}} \startdata

\bf NGC 2168 \rm & \\

A1 & 0 $\leq$ $R$ $<$ 5 & 0 $\leq$ $R$ $<$ 1456  \\

A2 & 5 $\leq$ $R$ $<$ 10 & 1456 $\leq$ $R$ $<$ 2912 \\

A3 & 10 $\leq$ $R$ $<$ 15 & 2912 $\leq$ $R$ $<$ 4369 \\

A4 & 15 $\leq$ $R$ $<$ 20 & 4369 $\leq$ $R$ $<$ 5825 \\

Global & 0 $\leq$ $R$ $<$ 20 & 0 $\leq$ $R$ $<$ 5825 \\

\bf NGC 2323 \rm & \\

A1 & 0 $\leq$ $R$ $<$ 3 & 0 $\leq$ $R$ $<$ 873  \\

A2 & 3 $\leq$ $R$ $<$ 7 & 873 $\leq$ $R$ $<$ 2039 \\

A3 & 7 $\leq$ $R$ $<$ 11 & 2039 $\leq$ $R$ $<$ 3204 \\

A4 & 11 $\leq$ $R$ $<$ 15 & 3204 $\leq$ $R$ $<$ 4368 \\

Global & 0 $\leq$ $R$ $<$ 15 & 0 $\leq$ $R$ $<$ 4368 \\

\enddata

\end{deluxetable}

\clearpage

\begin{deluxetable}{lll}
\tabletypesize{\scriptsize} \tablecaption{Summary of Results for NGC 2168 
\label{table4}} \tablewidth{0pt}
\tablehead{} \startdata
Position: & \\
$\alpha_{\rm J2000}$  & RA                 & = $06^{\rm h}08^{\rm m}54.0^{\rm s}$   \\ 
$\delta _{\rm J2000}$ & declination        & = $+24^{\rm o}20.0'$       \\
$l_{\rm J2000}$       & Galactic longitude & = $186.58^{\rm o}$             \\   
$b_{\rm J2000}$       & Galactic latitude  & = $2.18^{\rm o}$               \\ 
\\
Distance and Reddening: & \\
$(m {\rm-}M)_{V}$     & apparent distance modulus     & = 10.42   \\
$E(B {\rm-}V)$      & reddening                     & = 0.20    \\
$A_{V}$             & visual extinction             & = 0.62    \\
$(m {\rm-}M)_{\rm 0}$    & true distance modulus         & = 9.80    \\
$d$           & distance from Sun          & = 912 $\pm \ ^{70}_{65}$ pc  \\
$z$           & distance above Galactic plane                & = 36.6 pc   \\
\\
Age: & \\
$t_{\rm Dynamical}$   & dynamical relaxation timescale  & = 111 Myrs    \\
$t_{\rm Isochrone}$   & main-sequence turnoff age       & = 180 Myrs    \\
$t_{\rm WD}$          & white dwarf cooling age         & $\sim$ 190 Myrs \\
\\
Metallicity: & \\
$Z^{\rm 1}$             & heavy metal abundance        & = 0.012         \\
\\
Size: & \\
$D$       & linear diameter                         & = 10.6 pc       \\
$\Theta$  & angular diameter                        & = 40$'$         \\
$N$       & number of stars down to M$_{\rm V}$ = 15$^{2}$  & $\sim$ 1500    \\
\\
Stellar Distribution: & \\
$\alpha$      & mass function slope (Salpeter = 2.35) & = 2.29 $\pm$ 0.27  \\

\enddata

\tablenotetext{1}{These values were not spectroscopically determined.}
\tablenotetext{2}{Our study does not include the entire cluster.}

\end{deluxetable}


\clearpage

\begin{deluxetable}{lll}
\tabletypesize{\scriptsize} \tablecaption{Summary of Results for NGC 2323 
\label{table5}} \tablewidth{0pt}
\tablehead{} \startdata
Position: & \\
$\alpha_{\rm J2000}$  & RA                   & = $07^{\rm h}02^{\rm m}48.0^{\rm s}$  \\ 
$\delta _{\rm J2000}$ & declination          & = $-08^{\rm o}22'36''$          \\
$l_{\rm J2000}$       & Galactic longitude   & = $221.67^{\rm o}$              \\   
$b_{\rm J2000}$       & Galactic latitude    & = $-1.24^{\rm o}$               \\ 
\\
Distance and Reddening: & \\
$(m {\rm-}M)_{V}$   & apparent distance modulus   & = 10.68     \\
$E(B {\rm-}V)$      & reddening                   & = 0.22      \\
$A_{V}$       & visual extinction               & = 0.68      \\
$(m {\rm-}M)_{\rm 0}$   & true distance modulus       & = 10.0      \\
$d$           & distance from Sun         & = 1000 $\pm \ ^{81}_{75}$ pc  \\
$z$           & distance below Galactic plane                & = 22.7 pc  \\
\\
Age: & \\
$t_{\rm Dynamical}$   & dynamical relaxation timescale  & = 102 Myrs   \\
$t_{\rm Isochrone}$   & main-sequence turnoff age       & = 130 Myrs   \\
$t_{\rm WD}$          & white dwarf cooling age         & = -------------   \\
\\
Metallicity: & \\
$Z^{\rm 1}$             & heavy metal abundance         & = 0.020      \\
\\
Size: & \\
$D$       & linear diameter                         & = 8.72 pc    \\
$\Theta$  & angular diameter                        & = 30$'$      \\
$N$   & number of stars down to M$_{\rm V}$ = 15  & $\sim$ 3200  \\
\\
Stellar Distribution: & \\
$\alpha$      & mass function slope (Salpeter = 2.35)  & = 2.94 $\pm$ 0.15  \\

\enddata

\tablenotetext{1}{These values were not spectroscopically determined.}

\end{deluxetable}

\end{document}